\documentclass[runningheads]{llncs}
\usepackage{graphicx}
\usepackage[english]{babel}
\usepackage{amsfonts}
\usepackage{amssymb}
\usepackage{amsmath}
\usepackage{blindtext}
\usepackage{graphicx}
\usepackage{marvosym}
\usepackage{soul}
\usepackage[dvipsnames]{xcolor}
\usepackage{bbm}
\usepackage{siunitx}

\usepackage{mathtools}
\usepackage{caption}
\usepackage{subcaption}
\newcommand{\beq}{\begin{equation}}
\newcommand{\eeq}{\end{equation}}

\newcommand{\bmulti}{\begin{multline*}}
\newcommand{\emulti}{\end{multline*}}

\newcommand{\dd}{\mathrm{d}}

\newcommand{\EE}{{\mathbb E}}

\newcommand{\bgm}{\begin{bmatrix}}
\newcommand{\enm}{\end{bmatrix}}
\newcommand{\mult}{\begin{multline*}}
\newcommand{\emul}{\end{multline*}}

\newcommand{\cK}{{\cal K}}

\newcommand{\af}{\alpha}

\newcommand{\Gm}{\Gamma}

\newcommand{\gm}{\gamma}

\newcommand{\be}{\begin{equation}}

\newcommand{\ee}{\end{equation}}

\definecolor{nicegreen}{HTML}{198f0a}

\newcommand{\R}{\mathbb{R}}
\newcommand{\N}{\mathbb{N}}

\newcommand{\cF}{\mathcal{F}}
\newcommand{\cL}{\mathcal{L}}
\newcommand{\cM}{\mathcal{M}}

\newcommand{\cN}{\mathcal{N}}

\newcommand{\E}{\mathbb{E}}
\newcommand{\dist}{\text{dist}}

\renewcommand{\Re}{\mathbb{R}}

\makeatletter
\newcommand*\bigcdot{\mathpalette\bigcdot@{.5}}
\newcommand*\bigcdot@[2]{\mathbin{\vcenter{\hbox{\scalebox{#2}{$\m@th#1\bullet$}}}}}
\makeatother
\newcommand{\interior}[1]{%
  {\kern0pt#1}^{\mathrm{o}}%
}

\begin{document}
\title{Currents and $K$-functions for Fiber Point Processes}
%
%
\author{}
\institute{}
\iftrue
\author{Pernille EH.\  Hansen\inst{1}\orcidID{0000-0002-5171-5060} \and
Rasmus Waagepetersen\inst{2}\orcidID{0000-0001-6911-0089} \and Anne Marie Svane\inst{2}\orcidID{0000-0001-6356-0484}\and
Jon Sporring\inst{1}\orcidID{0000-0003-1261-6702} \and
Hans JT.\  Stephensen\inst{1}\orcidID{0000-0001-8245-0571}
Stine Hasselholt \inst{3}\orcidID{0000-0001-5362-6371} \and
Stefan Sommer\inst{1}\orcidID{0000-0001-6784-0328}
}

\authorrunning{ PEH.\ Hansen et al.} 
%
\institute{Department of Computer Science, University of Copenhagen, Copenhagen, Denmark
\email{\{pehh, sporring, hast, sommer\}@di.ku.dk}\\\and
Department of Mathematical Sciences, Aalborg University, Aalborg, Denmark\\
\email{\{rw, annemarie\}@math.aau.dk} \\\and Stereology and Microscopy, Aarhus University, Aarhus, Denmark \\
\email{stha@clin.au.dk}  }
\fi

\maketitle              
\begin{abstract}
Analysis of images of sets of fibers such as myelin sheaths or skeletal muscles must account for both the spatial distribution of fibers and differences in fiber shape. This necessitates a combination of point process and shape analysis methodology. In this paper, we develop a $K$-function for shape-valued point processes by embedding shapes as currents, thus equipping the point process domain with metric structure inherited from a reproducing kernel Hilbert space. We extend Ripley's $K$-function which measures deviations from spatial homogeneity of point processes to fiber data. The paper provides a theoretical account of the statistical foundation of the $K$-function and its extension to fiber data, and we test the developed $K$-function on simulated as well as real data sets. This includes a fiber data set consisting of myelin sheaths, visualizing the spatial and fiber shape behavior of myelin configurations at different debts.
\end{abstract}
%
%
%

\section{Introduction}
We present a generalization of Ripley's $K$-function for shape-valued point processes, in particular, for point processes where each observation is a curve in $\R^3$, a fiber. Fiber structures appear naturally in the human body, for example in tracts in the central nervous system and in skeletal muscles. The introduced $K$-function captures both spatial and shape clustering or repulsion, thus providing a powerful descriptive statistic for analysis of medical image of sets of fiber or more general shape data.
As an example, Fig. 1 displays myelin sheaths in four configurations from different debts in a mouse brain. We develop the methodology to quantify the visually apparent differences in both spatial and shape distribution of the fibers.

\subsection{Background}
Ripley's $K$-function \cite{ripley_second-order_1976} is a well-known tool for analyzing second order moment structure of point processes \cite{baddeley_spatial_2015} providing a measure of deviance from complete spatial randomness in point sets. For a stationary point process, $K(t)$ gives the expected number of points within distance $t$ from a typical point. An estimator of Ripley's $K$-function for a point set $\{p_i\}_{i=1}^n$ inside an observation window $W$ is,
\begin{align}\label{RipleysKfunction}
	\hat{K}(t) = \frac{1}{n\hat{\lambda}} \sum_{i\neq j }1[\dist(p_i,p_j)<t]
\end{align} 
where $\hat{\lambda} = \frac{n}{|W|}$ is the sample intensity, $|W|$ is the volume of the observation window, and $1$ is the indicator function. By comparing $\hat{K}(t)$ with the
$K$-function corresponding to complete spatial randomness, we can measure the deviation from spatial homogeneity. Smaller values of $\hat{K}(t)$ indicate clustering whereas the points tend to repel each other for greater values.\par
\begin{figure}
\centering
\begin{tabular}{cccc}
ST01 & ST06 & ST17 & ST20
\\\includegraphics[width=0.23\textwidth,trim={3.5cm 1.8cm 1.5cm 3.5cm}, clip]{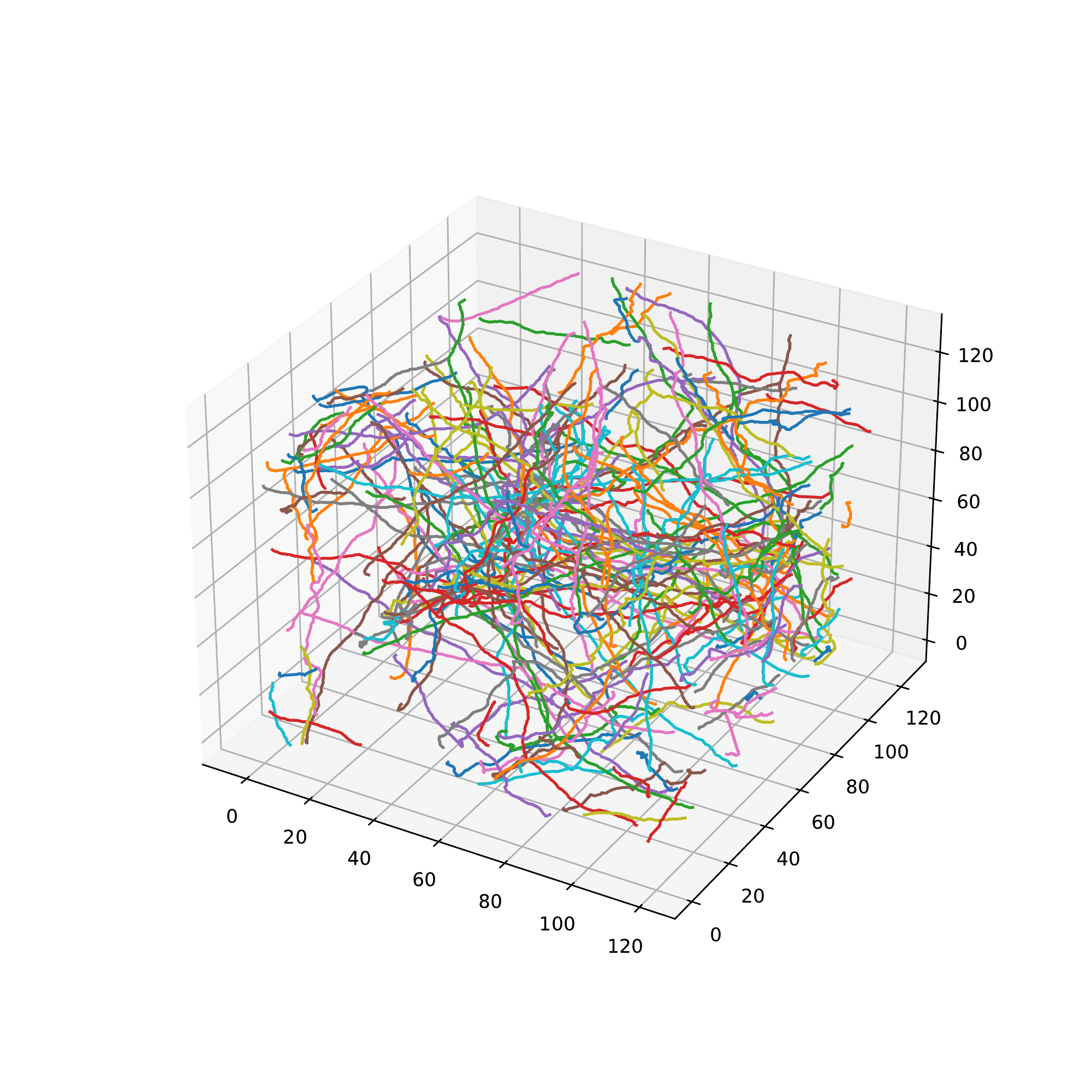}
&\includegraphics[width=0.23\textwidth,trim={3.5cm 1.8cm 1.5cm 3.5cm}, clip]{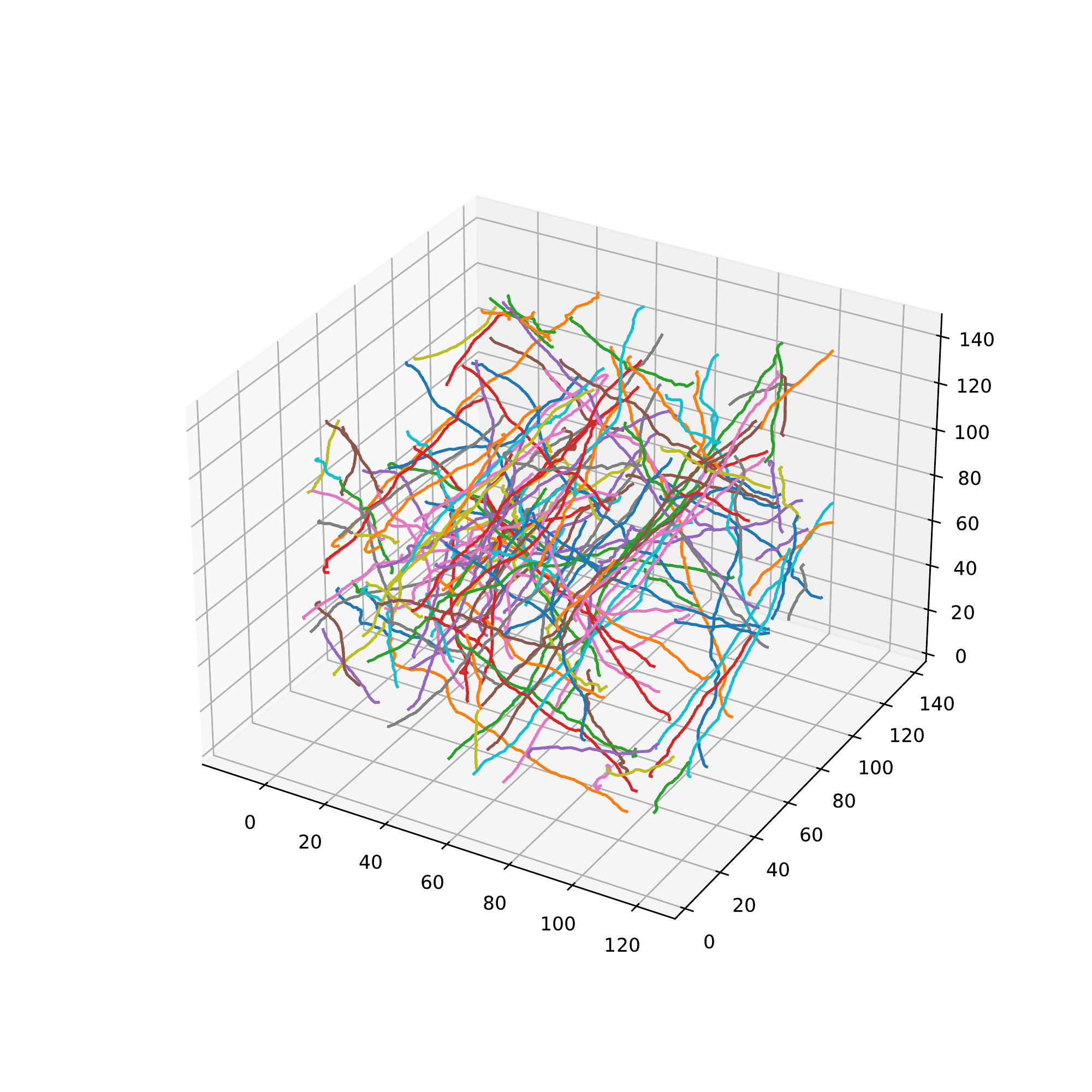}
&\includegraphics[width=0.23\textwidth,trim={3.5cm 1.8cm 1.5cm 3.5cm}, clip]{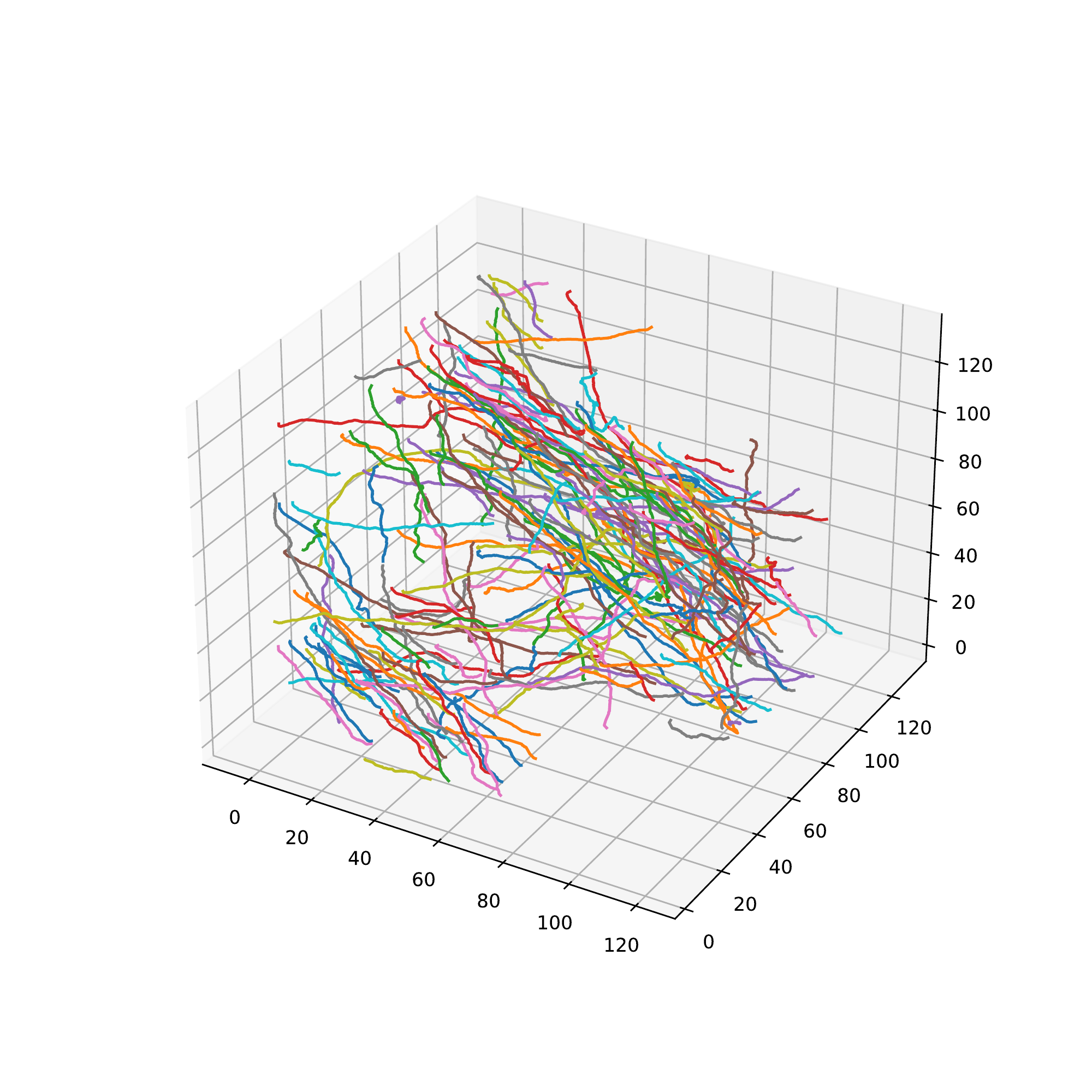}
&\includegraphics[width=0.23\textwidth,trim={3.5cm 1.8cm 1.5cm 3.5cm}, clip]{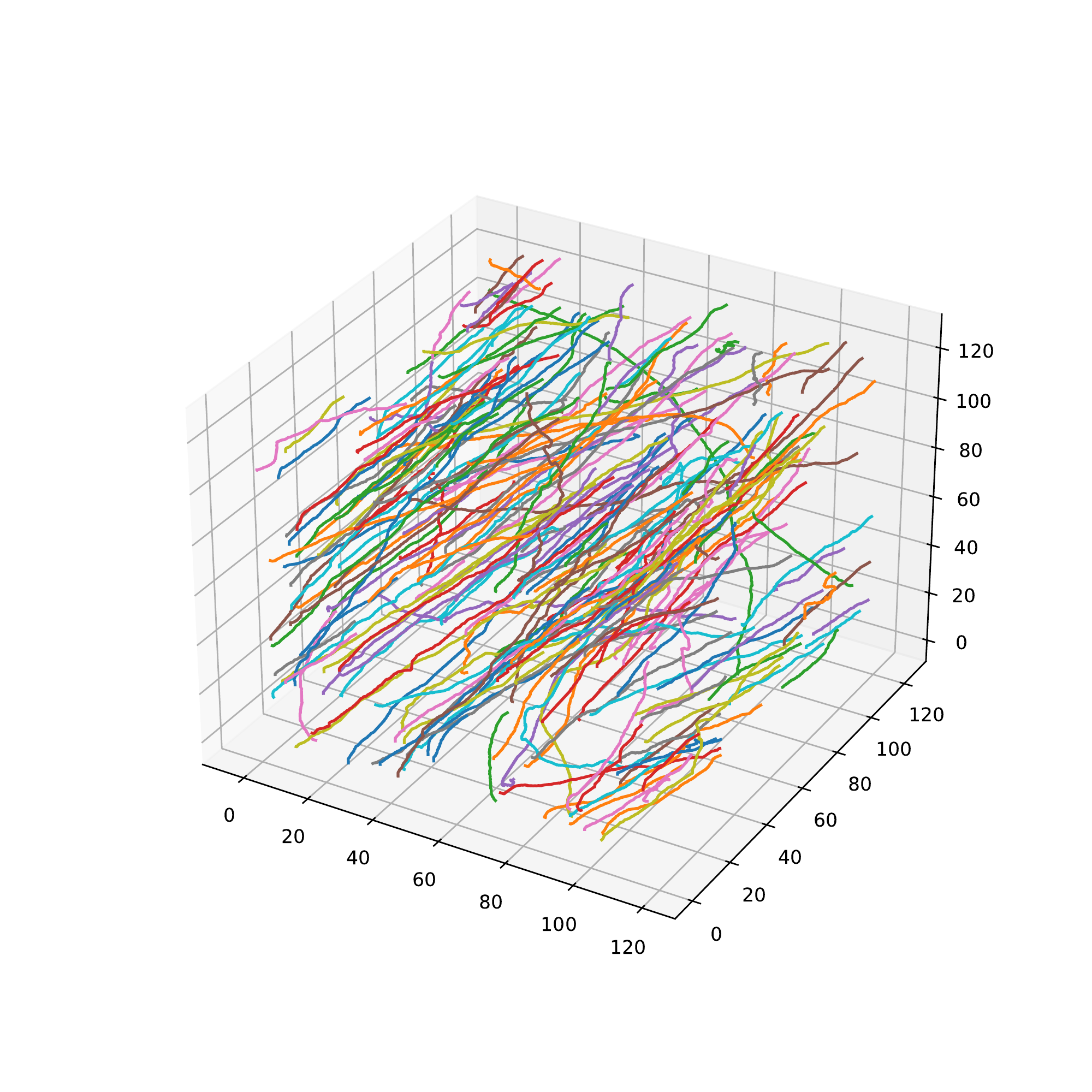}
\\{\includegraphics[width=0.23\textwidth, trim={ 1.2cm 0cm 10.45cm 1.2cm}, clip]{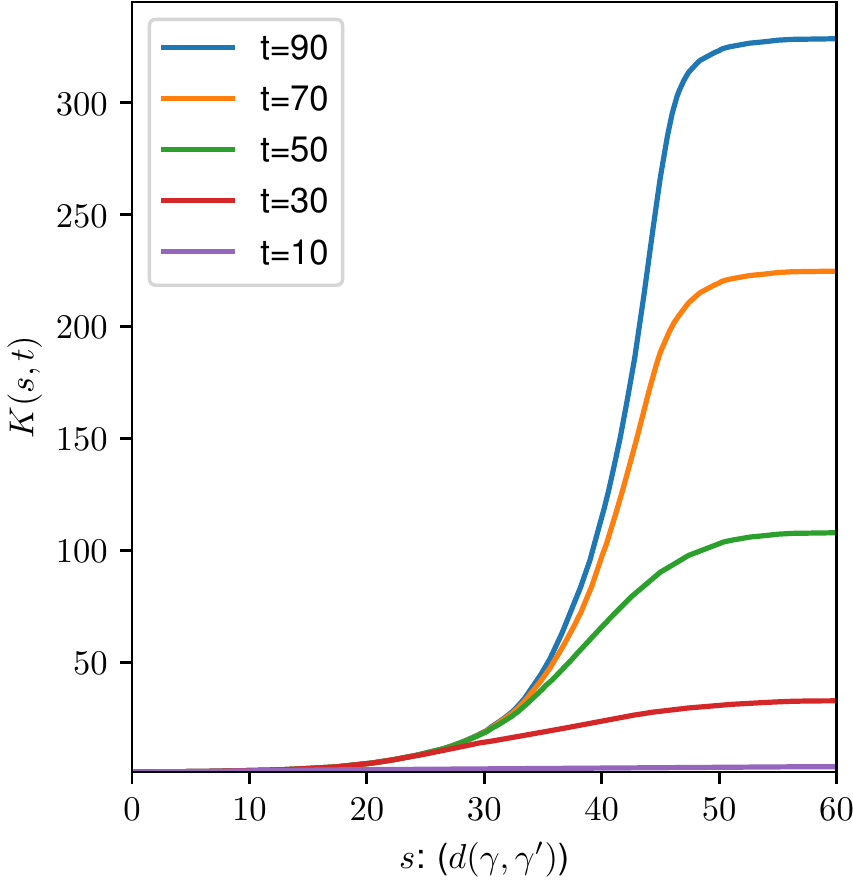}}
&{\includegraphics[width=0.23\textwidth, trim={ 1.2cm 0cm 10.45cm 1.2cm}, clip]{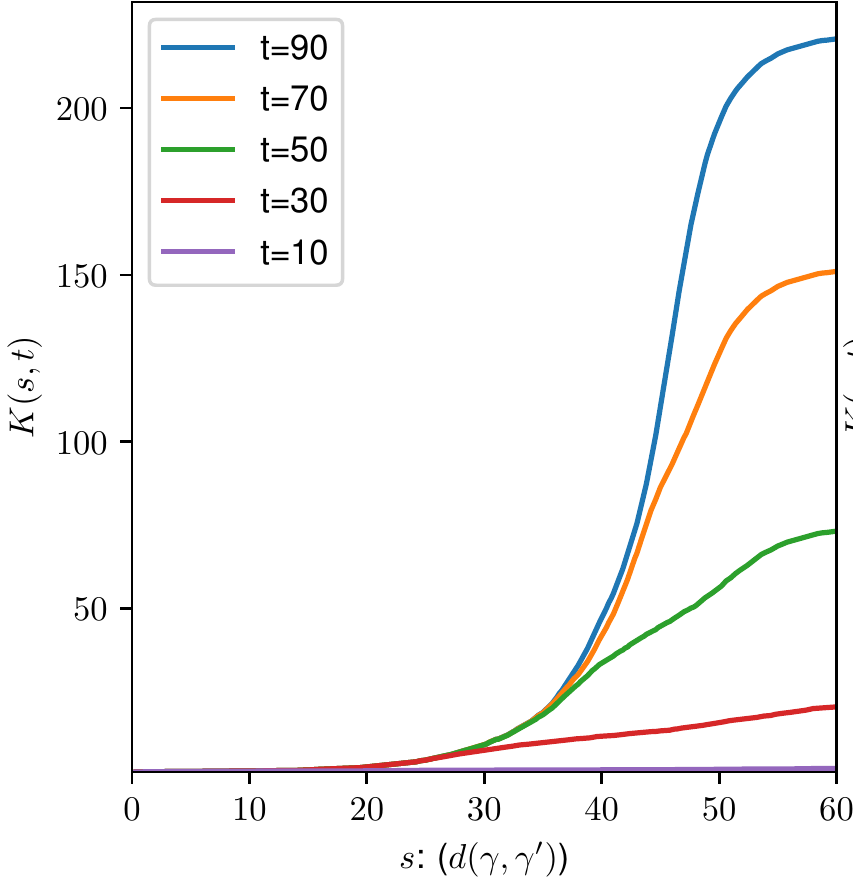}}
&{\includegraphics[width=0.23\textwidth, trim={ 1.2cm 0cm 10.45cm 1.2cm}, clip]{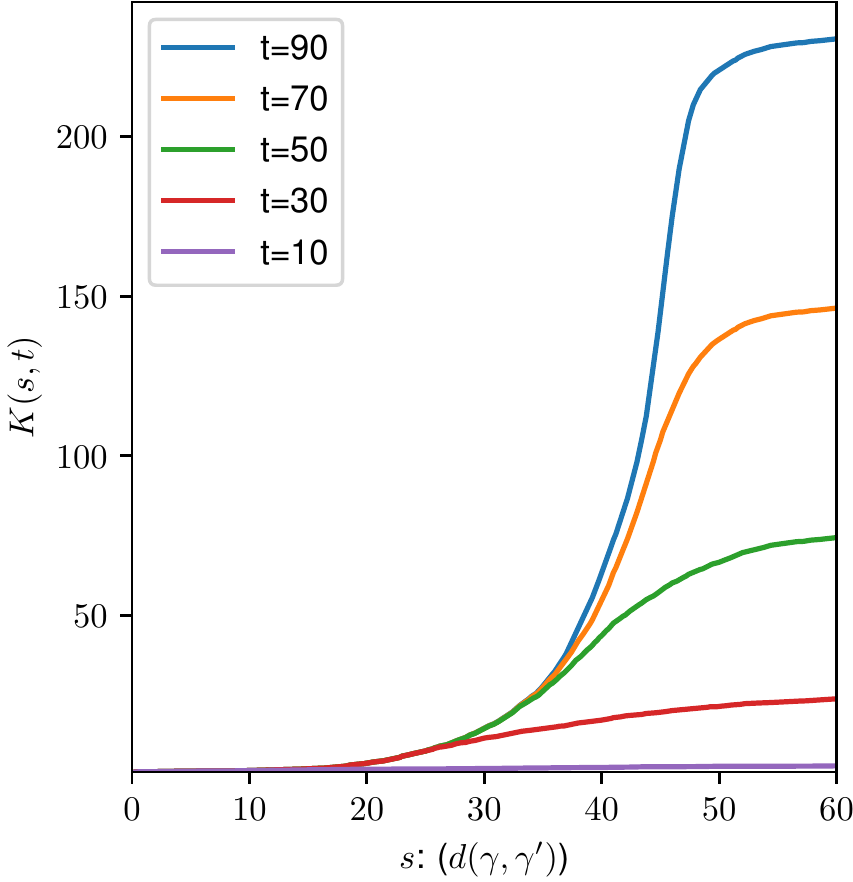}}
&{\includegraphics[width=0.23\textwidth, trim={ 1.2cm 0cm 10.45cm 1.2cm}, clip]{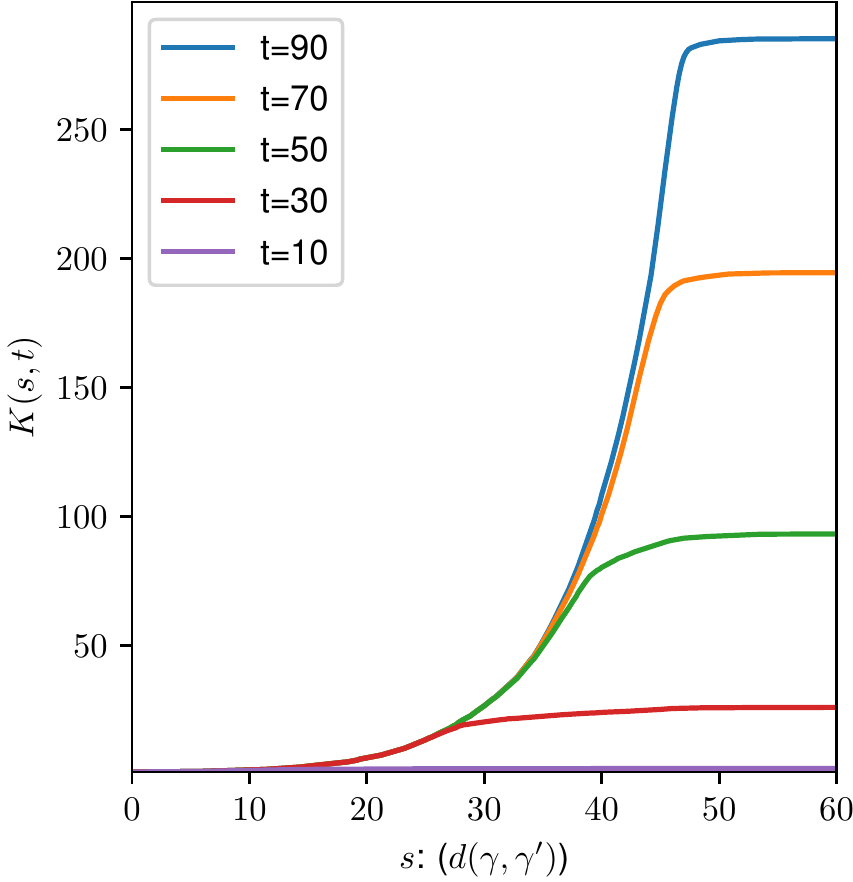}}
\\\includegraphics[width=0.23\textwidth, trim={ 9.85cm 0cm 1.8cm 1.2cm}, clip]{plots/K_ST01.pdf}
&\includegraphics[width=0.23\textwidth, trim={ 9.85cm 0cm 1.8cm 1.2cm}, clip]{plots/K_ST06.pdf}
&\includegraphics[width=0.23\textwidth, trim={ 9.85cm 0cm 1.8cm 1.2cm}, clip]{plots/K_ST17.pdf}
&\includegraphics[width=0.23\textwidth, trim={ 9.85cm 0cm 1.8cm 1.2cm}, clip]{plots/K_ST20.pdf}
\end{tabular}
\caption{$K$-functions for samples of myelin sheaths. Each column corresponds to measured on a data set. (Top row): the centerlines of the myelin sheathed axons. (Middle row): The $K$-function for fixed values of $t$. (Bottom row): The $K$-function for fixed values of $s$.}
\label{myelin3d}
\end{figure}
Generalizations of Ripley's $K$-function have previously been considered for curve pieces in \cite{chiu_stochastic_2013} and several approaches were presented in \cite{sporring_generalizations_2019} for space curves. In this paper, we present a $K$-function inspired by the currents approach from \cite{sporring_generalizations_2019} and provide the theoretical account for the statistical foundation.
\par 
The challenges when generalizing the $K$-function to shape-valued point processes arise in defining a distance measure on the shape space and determining a meaningful descriptive quantity that is well-defined and that we are able to estimate. In this paper, we provide a well-defined $K$-function for point processes in general metric spaces and use the embedding of shapes as currents to obtain a distance measure of shapes. 
\subsection{Contributions and outline}
We construct the following two-parameter $K$-function for a curve-valued point process $X$
\begin{equation}  \hat{K}(t,s) = \frac{1}{|W|\lambda}
\sum_{\gm \in X: c(\gm) \in W} \sum_{\gm' \in X \setminus \{\gm\} }
1[\|c(\gm)-c(\gm')\| \le t, d_m(\gm,\gm') \le s]
\end{equation} 
for $t,s>0$, where $c(\gamma)$ is the center point of the curve $\gamma$, $d_m$ the minimal currents distance with respect to translation, and $\lambda$ is the spatial intensity of the center points. By introducing a second distance parameter, we are able to separate the spatial distance of the curves from the difference in shape, allowing us to measure both spatial and shape homogeneity.
\par
The paper thereby presents the following contributions: 
\begin{enumerate}
    \item A $K$-function for shape-valued point processes along with a theoretical account for the statistical foundation.
    \item We suggest a certain fiber process which we argue corresponds to complete randomness of points,  an analogue of the Poisson process.
    \item An application of the $K$-function to several generated data set and a real data set of myelin sheaths.
\end{enumerate}
\section{Shapes as currents}
\subsection{Shape-valued point processes}
We model a random collection of shapes as a point process on the space of shapes. Shape spaces are usually defined as the space of embeddings $B_e(\cM,\R^d)$ of a manifold $\cM$ into $\R^d$ \cite{bauer_overview_2014}. For example, the space of closed curves in $\R^3$ is $B_e(S^1,\R^3)$, where $S^1$ denotes the 1-sphere, and the space of fibers is $B_e(I,\R^3)$ for some real interval $I$.
\par
Formally, a point process $X$ on a metric space $S$ is a measurable map from some probability space $(\Omega,\cF,P)$ into the space of locally finite subsets of $S$. Thus, for each $\omega\in \Omega$, $X(\omega) \subseteq S$, and for every compact Borel set $B\subseteq S$, $X(\omega) \cap B$ is a finite set.  Measurability of $X$ means that all sets of the form $\{ \omega \in \Omega | \#(X(\omega)\cap B)=m \}$, where $m\in \N_0$ and $B\subseteq S$ is a Borel set, must be measurable. 
\par
There are different ways to endow $B_e(\cM,\R^d)$ with a metric \cite{michor_metric_2008}.  In this paper, we consider the representation of shapes as currents embedded in the dual space of a reproducing kernel Hilbert space (RKHS). Thus, the RKHS metric induces a metric for our shape space. This can be combined with the Euclidean metric on $\R^d$ to obtain a suitable metric on $B_e(\cM,\R^d)$. This approach is very useful due to its generality and computability, as it requires very little information about the shape.

\subsection{Shapes as currents}
Shapes are usually more difficult to work with than points, as they usually cannot be captured in any finite dimensional vector space. An approach already considered for anatomical structures  \cite{durrleman_statistical_2009} \cite{vaillant_surface_2005} is embedding shapes as currents.
We will give a brief introduction to this setup and refer to \cite{durrleman_statistical_2009} for a detailed description.
We can characterize  a piece-wise smooth curve  $\gamma\in B_e(I,\R^d)$  by computing its path-integral of all vector fields $w$ 
 \begin{align}\label{Vgamme}
 V_\gamma(w) = \int_\gamma w(x)^t\tau(x)d\lambda(x),
 \end{align}
 where $\tau(x)$ is the unit tangent of $\gamma$ at $x$ and $\lambda$ is the length measure on the curve. Likewise, an oriented hypersurface $S$ embedded in $\R^d$ can be characterized by its flux integral of vector fields $w$ 
 \begin{align}
 V_S(w) = \int_S w(x)^tn(x)d\lambda(x),
 \end{align}
 where $n(x)$ is the unit normal at $x$ and $\lambda$ is the  surface area measure on $S$. These are both examples of representing shapes as currents, i.e.\ as elements in the dual space of the space of vector fields on $\R^d$. Formally, the space of $m$-currents $C_m$ is the dual space of the space $C^0(\R^d, (\Lambda^m\R^d)^*)$ of differential $m$-forms. 
 
It is not only curves and hypersurfaces that can be represented as currents. Let $\cM$ be an oriented rectifiable sub-manifold of dimension $m$ in $\R^d$ with positively oriented basis of the tangent space $u_1(x),...,u_m(x)$ for all $x\in \cM$. The sub-manifold $\cM$ can be embedded into the space of $m$-currents as the current 
\begin{equation}\label{generalembedding}
    T_\cM(w) = \int_\cM I(x) w(x) \Big( \frac{u_1(x) \wedge ... \wedge u_m }{|u_1(x) \wedge ... \wedge u_m|} \Big) d\lambda(x)
\end{equation}
where $w\in C^0(\R^d, (\Lambda^m\R^d)^*)$ is an $m$-differential form and $I:T\to \R$ is a scalar function satisfying $\int_T |I(x)| d\lambda(x)<\infty$ \cite{durrleman_statistical_2009}. Since shapes are embedded sub-manifolds, this means that shapes can be embedded into $C_m$.

\subsection{Reproducing kernel Hilbert space metric on shapes}
The space of $m$-currents $C_m$ is continuously embedded into the dual space of a reproducing kernel Hilbert space (RKHS) $H$ with arbitrary kernel $K_H:\R^d\times\R^d \to \R^{d\times d}$ \cite{durrleman_statistical_2009}.
It follows from Riesz representation theorem that $v\in H$ can be embedded in the dual space $H^*$ as the functional $\cL_H(v)\in H^*$ defined by $\cL_H(v)(w)=  \langle v,w\rangle_H$ for $w\in H$.
\par
Elements $v(y) = K_H(x,y)\alpha$ form a basis for $H$ where $x,\alpha\in \R^d$, and basis elements in $H$ are lifted to basis elements in $H^*$ as $\delta_x^\alpha := \cL_H(v)$ which are called the Dirac delta currents.
The element $V_\gamma$ from \eqref{Vgamme} can be written in terms of the basis elements $\delta^{\tau_{\gamma}(x_i)}_{x_i}$ where $\tau_{\gamma}(x_i)$ are the unit tangent vectors of $\gamma$ at $x_i$.  This means that the curve $\gamma$ is embedded into $H^*$ as the $1$-current \begin{equation}
	V_\gamma(w) = \int_\gamma w(x)^t\tau_{\gamma}(x)d\lambda(x) = \int_\gamma \delta_{x}^{\tau_{\gamma}(x)}(w)d\lambda(x)
	\label{oneCurrent}
\end{equation}
where $\lambda$ is the length measure on the curve. Furthermore, it is approximated by the Riemann sum of Dirac delta currents $V_\gamma(w) \approx \tilde{V}_\gamma(w) =\sum_i \delta_{x_i}^{\tau(x_i)\Delta x_i}(w)$ where $x_i$ are sampled points along the curve according to $\lambda$.
The dual space $H^*$ inherits the inner product from the inner product on the RKHS via the inverse mapping $\cL_H^{-1}$, so that the inner product for two curves $\gamma_1$ and $\gamma_2$ in $H^*$ is
\begin{align}
	\langle V_{\gamma_1}, V_{\gamma_2}\rangle_{H^*} = \int_{\gamma_1}\int_{\gamma_2} \tau_{\gamma_1}^t(x)K_H(x,y)\tau_{\gamma_2}(y)d\lambda_{\gamma_2}(x)d\lambda_{\gamma_1}(y).
\end{align} 
Writing $||V_{\gamma}||^2_{H^*}=\langle V_{\gamma}, V_{\gamma}\rangle_{H^*}$, we finally arrive at the currents distance of two curves $\gamma_1$ and $\gamma_2$
\begin{align}
	d_c(V_{\gamma_1}, V_{\gamma_2}) =||V_{\gamma_1}- V_{\gamma_2}||_{H^*} = \Big( ||V_{\gamma_1}||_{H^*}^2 + ||V_{\gamma_2}||_{H^*}^2 -2\langle V_{\gamma_1}, V_{\gamma_2}\rangle_{H^*} \Big)^{1/2}.
\end{align}
In practice, we usually don't know the orientation of the curves, thus we choose to consider the minimal distance between them,
\begin{equation}
	d(V_{\gamma_1}, V_{\gamma_2}) = \min\{ d_c(V_{\gamma_1}, V_{\gamma_2}), d_c(V_{\gamma_1}, V_{-\gamma_2})  \}
\label{mindistance}
\end{equation}
where $-\gamma_2$ denotes the curve with opposite orientation of $\gamma_2$. If the orientation of the data is important, this step may be omitted. From \eqref{oneCurrent} we see that $K_H$ serves as a weight of the inner product between $\tau_1(x_i)$ and $\tau_2(y_j)$ depending on the positions $x_i$ and $y_j$. 

\subsection{A note on short lines and generalized Gaussian kernels}
To illustrate the distance metric, consider the generalized Gaussian kernel 
\begin{equation}\label{gauskernel}
    K^p_\sigma(x,y) = \exp\Big({\frac{-|x-y|^p}{2\sigma^p}}\Big)\text{Id},
\end{equation} 
where $\sigma , p\in (0,\infty] $, and consider two lines of equal length parametrized by $l_u(t) = x_u + u t,\, l_v(t)=x_v+v t$ where $x_u,x_v,u,v \in \Re^d$ and $0<t\leq T\in\Re$. For very short lines far from each other, i.e., $T/|x_u-x_v|\rightarrow 0$, we have,
\begin{equation}
\frac{d(\tilde{V}_{l_u},\tilde{V}_{l_v})^2}{{T^2}} \rightarrow 
d_0 -2\exp\big({\frac{-|x_u-x_v|^p}{2\sigma^p}}\big)d_1,
\end{equation}
where $d_0 = u^tu + v^tv$ and $d_1=\max\left(u^tv,-u^tv\right)$. Since the $d_0$ and $d_1$ are constants, and since the exponential and the square root functions are both monotonic, then in the limit, $d(\tilde{V}_{l_u},\tilde{V}_{l_v})/T$ is one-to-one with $|x_u-x_v|^p$ which is one-to-one with $|x_u-x_v|$. Thus, for very short lines, $d/T$ is one-to-one with the euclidean distance between the points $x_u$, and $x_v$. Further, in the limit $p\rightarrow\infty$ we have,
\begin{equation}
\frac{d(\tilde{V}_{l_u},\tilde{V}_{l_v})^2}{T^2}\rightarrow 
\begin{cases}
d_0 -2\,d_1, &\text{ when } |x_u-x_v| < \sigma,\\
d_0 -2\exp\left(-\frac12\right)d_1, &\text{ when } |x_u-x_v| = \sigma,\\
d_0, &\text{ otherwise}.
\end{cases}
\end{equation}
Thus, for very short lines and very large exponents, $(d_0 - d^2/T^2)/(2d_1)$ converges to a unit step function in $|x_u-x_v|$ where the step is at $\sigma$.

\section{The $K$-function}
\subsection{Statistical Setup}
Let $S$ be the image of the embedding of $B_e(I,\R^d)$ into $C_1$. For brevity, $\gamma$ is identified with its representation in $C_1$. We model a random collection of curves as a point process $X$ in $S$. 
\par
Let $c:S\to \R^d$ be a center function on the space of fibers in $\R^d$ that associates a center point to each fiber. A center function should be translation covariant in the sense that $c(\gamma + x) = c(\gamma) + x$ for all $x\in \R^d$. It could be the center of mass or the midpoint of the curve with respect to curve length.
Let $S_c$ denote the space of centered fibers wrt. $c$, i.e., those $\gm\in S$ for which $c(\gamma)=0$. We define $\gamma_c := \gamma -c(\gamma)\in S_c$ to be the centering of $\gamma$.
\par
For Borel sets $B_1,A_1 \subset \R^d$ and $B_2,A_2 \subset S_c$, define the first moment measure
\[ \mu(B_1 \times B_2) =\EE \sum_{\gm \in X}1[ c(\gm) \in B_1, \gm_c \in B_2] \]
and the second moment measure
\[ \af((A_1 \times A_2) \times (B_1 \times B_2)) = \EE \sum_{\gm ,\gm' \in X}^{\neq} 1[ c(\gm) \in A_1, \gm_c \in A_2]1[ c(\gm') \in B_1, \gm'_c \in B_2]. \]
We assume that $\mu$ is translation invariant in its first argument, i.e.\
\[ \mu(B_1 \times B_2)= \mu((B_1+h)\times B_2) \]
for any $h \in \R^d$. This is for instance the case if the distribution of $X$ is invariant under translations. This implies that $\mu(\cdot \times B_2)$ is proportional to the Lebesgue measure for all $B_2$. Thus we can write
\[ \mu(B_1 \times B_2) = |B_1| \nu(B_2) \]
for some measure $\nu(\cdot)$ on $S_c$. Note that the total measure $\nu(S_c)$ is the spatial intensity of the center points, i.e.\ the expected number of center points in a unit volume window. In applications, this will typically be finite. In this case, we may normalize $\nu$ to obtain a probability measure which could be interpreted as the distribution of a single centered fiber.
\par
We define the reduced Campbell measure
\[ C^! (A_1\times A_2 \times F)= \EE \sum_{\gm \in X}1[ c(\gm) \in A_1, \gm_c \in A_2 , X \setminus \{\gm \} \in F] \]
where the "$!$" represents the removal of the point $\gamma$ from $X$. By disintegration,
\[  C^! (A_1\times A_2 \times F)= \int_{A_1 \times A_2} P^!_{c,\gm_c}(F) \mu(\dd (c,\gm_c)) .\]
By the standard proof, we get for any measurable function $h:\R^d \times S_c \times \cN \to [0,\infty)$
\begin{equation}\label{eq:cm} 
\EE \sum_{\gm \in X}h(c(\gm),\gm_c,X \setminus \{\gm \}) = \int_{\R^d \times S_c} \EE^!_{c,\gm_c}h(c,\gm_c,X) \mu(\dd (c,\gm_c)). 
\end{equation}
In particular,
\begin{equation}
\af((A_1 \times A_2) \times (B_1 \times B_2)) = \int_{A_1 \times A_2} \EE^!_{c,\gm_c} \sum_{\gm' \in X} 1[ c(\gm') \in B_1, \gm'_c \in B_2] \mu(\dd (c,\gm_c)).
\end{equation} 
Assume also that $\af$ is invariant under joint translation of  the arguments $A_1,B_1$. Then
\begin{align}{\cal K}_{c,\gm_c}(B_1 \times B_2) &:= \EE^!_{c,\gm_c} \sum_{\gm' \in X} 1[ c(\gm') \in B_1, \gm_c' \in B_2] \\
&\phantom{:}= \EE^!_{0,\gm_0} \sum_{\gm' \in X} 1[ c(\gm') \in B_1-c, \gm_c' \in B_2] \\
&\phantom{:}= {\cal K}_{0,\gm_0}((B_1-c) \times B_2) 
.\end{align}
Assume that also $\EE^!_{c,\gm_c}h(c,\gm_c,X)$ does not depend on $c$, which is true if the distribution of $X$ is invariant over translations. Then, using \eqref{eq:cm} and the factorization of $\mu$,
\begin{align} \nonumber \EE \sum_{\gm \in X } 1[c(\gm) \in W]h(c(\gm),\gm_c,X \setminus \{\gm \}) &= \int_{W\times S_c} \EE^!_{c,\gm_c}h(c,\gm_c,X) \mu(\dd (c,\gm_c))\\ =  \int_{W\times S_c} \EE^!_{0,\gm_0}h(0,\gm_0,X) \mu(\dd (c,\gm_0)) &= |W| \int_{S_c} \EE^!_{0,\gm_0}h(0,\gm_0,X) \nu(\dd \gm_0) 
\end{align}
From this it follows that 
\[ E_h = \frac{1}{|W|} \sum_{\gm \in X} 1[c(\gm) \in W] h(c(\gm),\gm_c,X \setminus \{\gm \}) \]
is an unbiased estimator of
$ \int_{S_c} \EE^!_{0,\gm_0}h(0,\gm_0,X) \nu(\dd \gm_0).$
Furthermore, if $\nu(S_c)$ is finite,
\[ \int_{S_c} \EE^!_{0,\gm_0}h(0,\gm_0,X) \nu(\dd \gm_0)  = \nu(S_c) \EE_{\tilde{\nu}} \EE^!_{0,\Gm_0}h(0,\Gm_0,X)\]
where $\Gm_0$ is a random centered fiber with distribution $\tilde{\nu}(\cdot)=\nu(\cdot)/\nu(S_c)$ and $\EE_{\tilde{\nu}}$ is expectation with respect to this distribution of $\Gamma_0$. 
\subsection{$K$-function for fibers}
In order to define a $K$-function, we must make an appropriate choice of $h$. A seemingly natural choice for $h$ that coincides with \cite{sporring_generalizations_2019}, is 
\[ h(c,\gm_c,X) = \sum_{\gm' \in X} 1[d(\gm_c+c,\gm') \le t] =  \sum_{\gm' \in X} 1[d(\gm,\gm') \le t]. \]
However this choice allows the $K$-function to be a.s.\ infinite, due to the fact that $d(\gm,\gm')\leq \sqrt{2(||\gm||_{H^*}^2+||\gm'||_{H^*}^2)}$. If every curve in $X$ has $||\gamma||_{H^*}\leq M$, e.g.\ if the length of fibers is bounded, then choosing $t\geq 2M$ results in any fiber in $X$ having infinitely many neighbors within distance $t$. 
\par 
A solution is to separate the spatial distance of the curves from the difference in shape by introducing another radius parameter for the distance between center points. Accounting for spatial distance with this parameter,  we choose to minimize the influence of spatial distance by measuring the currents distance between the centered curves. 
Thus, we choose $h$ as
\begin{equation}
	h(c,\gm_c,X) = \sum_{\gm' \in X\setminus \{\gm\}}1[\|c(\gm)-c(\gm')\| \le t, d(\gm_c,\gm_c') \le s]
\end{equation}
for $s,t>0$, where $||c(\gm)-c(\gm')||$ is the usual distance in $\R^d$ between center points.
Thus we define the empirical $K$-function for $t,s>0$ as
\begin{equation}\label{eq:Kfix}  \hat{K}(t,s) = \frac{1}{|W|\nu(S_c)}
\sum_{\substack{\gm \in X: c(\gm) \in W, \\ \gm' \in X \setminus \{\gm\} }}
1[\|c(\gm)-c(\gm')\| \le t, d(\gm,\gm') \le s] .\end{equation}
Since $\nu(S_c) $ is the intensity of fiber centers, it is estimated by $N/|W|$ where $N$ is the observed number of centers $c(\gm)$ inside $W$. The $K$-function is the expectation of the empirical $K$-function  
\[K(t,s)=\E \hat{K}(t,s) =\EE_{\tilde{\nu}} \EE^!_{0,\Gm_0}h(0,\Gm_0,X) = \EE_{\tilde{\nu}} \cK_{0,\Gamma_0}(B(0,t)\times B_c(\Gamma_0,s)) \]
where $B(x,r) = \{y: ||x-y||\le r \}$ and $B_c(\gamma,s) = \{\gamma' : d(\gamma,\gamma')\le s\}$. 

\subsection{$K$-function for general shapes}
The currents metric and the $K$-function easily extends to shape-valued point processes with values in $B_e(\mathcal M, \Omega)$ for more general manifolds $\mathcal M$ and $\Omega\subset \R^d$. Shapes $\mathcal A, \mathcal B\in B_e(\mathcal M,\R^d)$ are embedded as $m$-currents $V_{\mathcal A}$ and $V_{\mathcal B}$ as in \eqref{generalembedding}. Since $C_m$ is continuously embedded into the dual RKHS $H^*$, we get the distance measure $d_c$ between shapes. 
%
\par
If $c: B_e(\mathcal M,\Omega)\to \R^d$ is a center function, then we can generalize the $K$-function to a point process $X$ with values in $B_e(\mathcal M,\Omega)$. Identifying elements of $B_e(\mathcal M, \Omega)$ with their embedding in $H^*$, we can write the same $K$-function
\begin{equation}
\hat{K}(t,s) = \frac{1}{|W|\lambda}
\sum_{\substack{\mathcal U \in X: c(\mathcal U) \in W, \\ \mathcal U' \in X \setminus \{\mathcal U\}} }
1[\|c(\mathcal U)-c(\mathcal U')\| \le t, d_m(\mathcal U,\mathcal U') \le s]
\end{equation}
for $t,s>0$, where $d_m$ is constructed as in Section $3.2$ and $\lambda$ is the spatial intensity of the center points.
\section{Experiments}
To obtain a measure of spatial homogeneity, Ripley's $K$-function for points is usually compared with the $K$-function for a Poisson process, $K_P(t) = \text{vol}(B_d(t))$, corresponding to complete spatial randomness.
We are now in a more complicated situation where the $K$-function has two parameters and we do not have a notion of complete randomness of fibers. 
The aim of the experiments on generated data sets is to analyze the behavior of the $K$-function on different types of distributions and suggest 
a fiber process that corresponds to complete randomness. This will serve as a way to compare the results in 4.2. 
\subsection{Generated data sets}
\begin{figure}
\centering
\begin{tabular}{cccc}
Uniform Lines  & Uniform Spirals & Uniform Brownian & Clustered Lines
\\{\includegraphics[width=0.23\textwidth,trim={3.5cm 2.8cm 2cm 3.5cm}, clip]{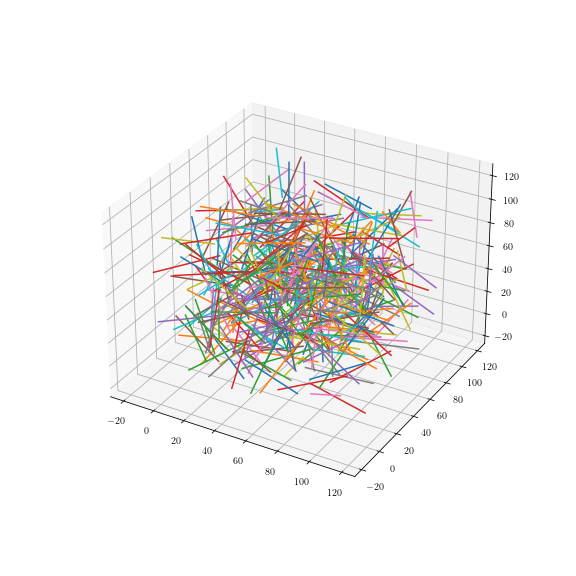}}
&{\includegraphics[width=0.23\textwidth,trim={3.5cm 2.8cm 2cm 3.5cm}, clip]{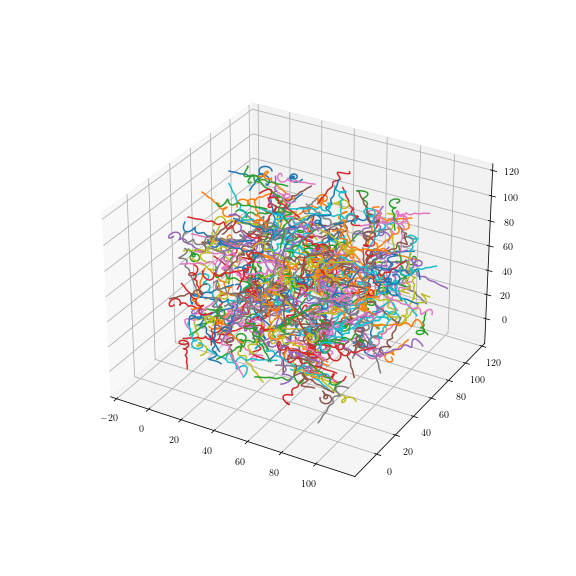}}
&{\includegraphics[width=0.23\textwidth,trim={3.5cm 2.8cm 2cm 3.5cm}, clip]{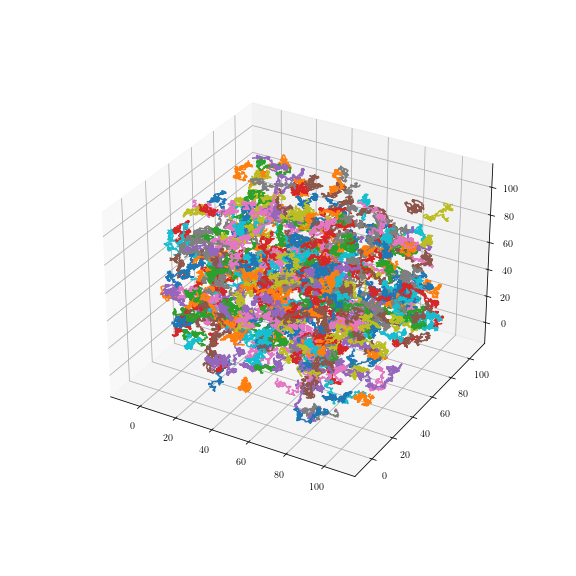}}
&{\includegraphics[width=0.23\textwidth,trim={3.5cm 2.8cm 2cm 3.5cm}, clip]{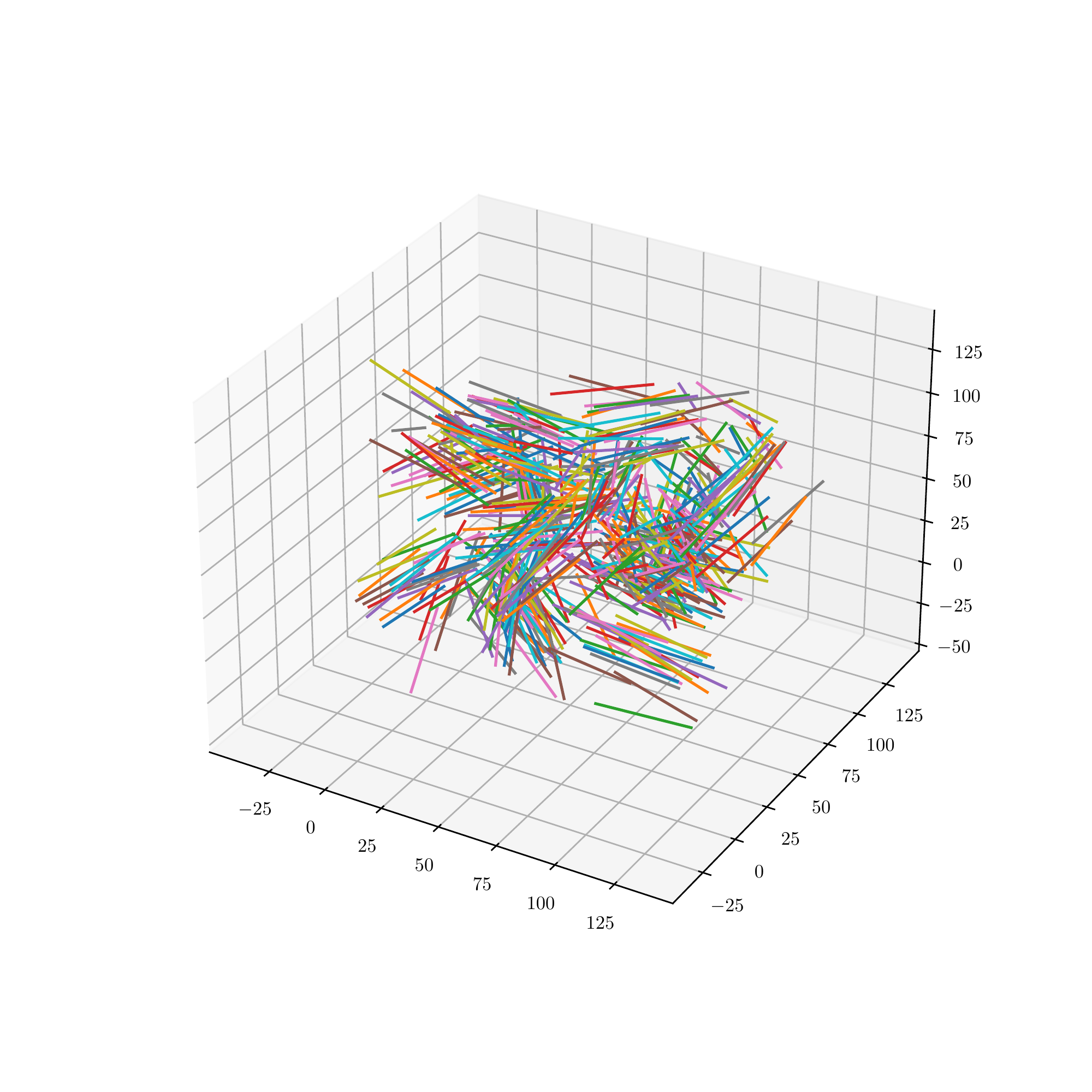}}
\\{\includegraphics[width=0.23\textwidth, trim={ 1.2cm .2cm 10.45cm 1.2cm}, clip]{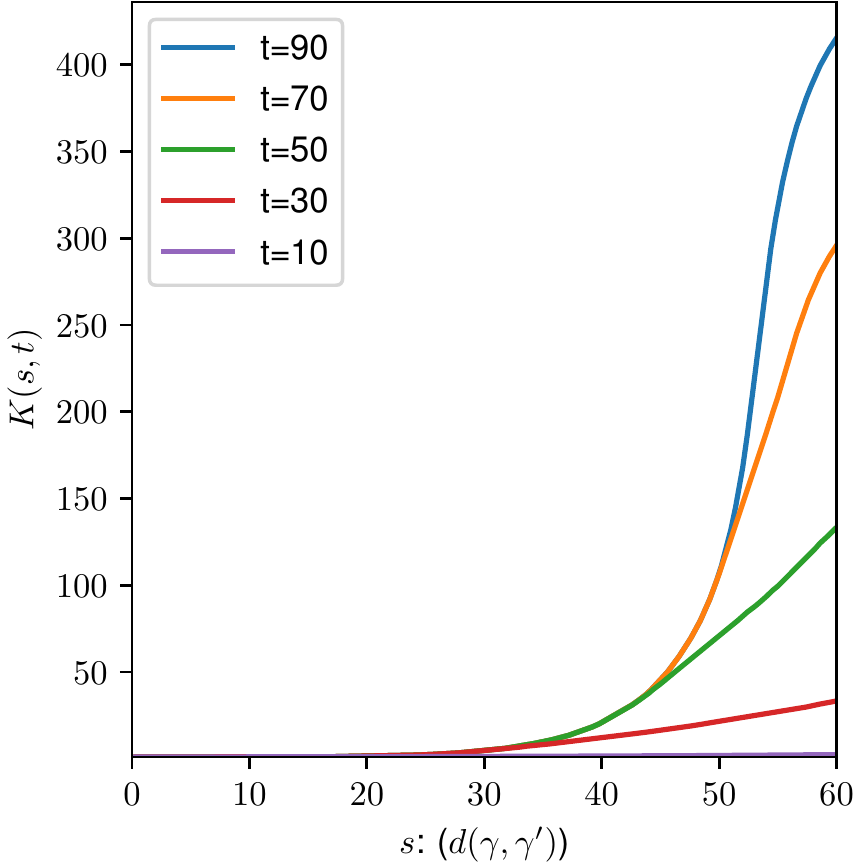}}
&{\includegraphics[width=0.23\textwidth, trim={ 1.2cm .2cm 10.45cm 1.2cm}, clip]{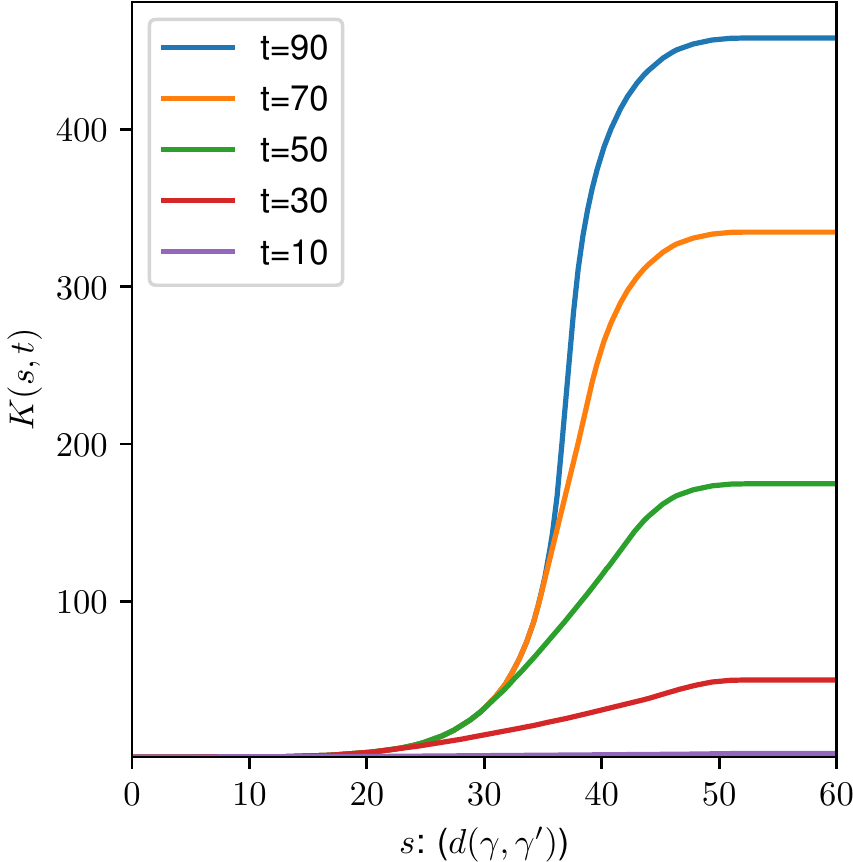}}
&{\includegraphics[width=0.23\textwidth, trim={ 1.2cm .2cm 10.45cm 1.2cm}, clip]{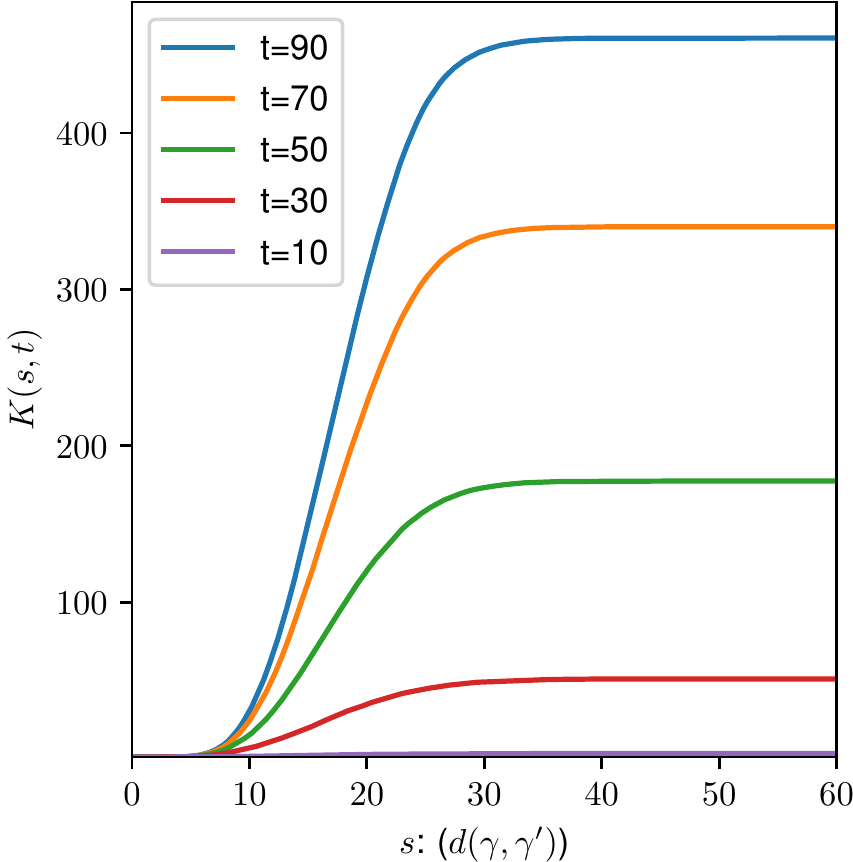}}
&{\includegraphics[width=0.23\textwidth, trim={ 1.2cm .2cm 10.45cm 1.2cm}, clip]{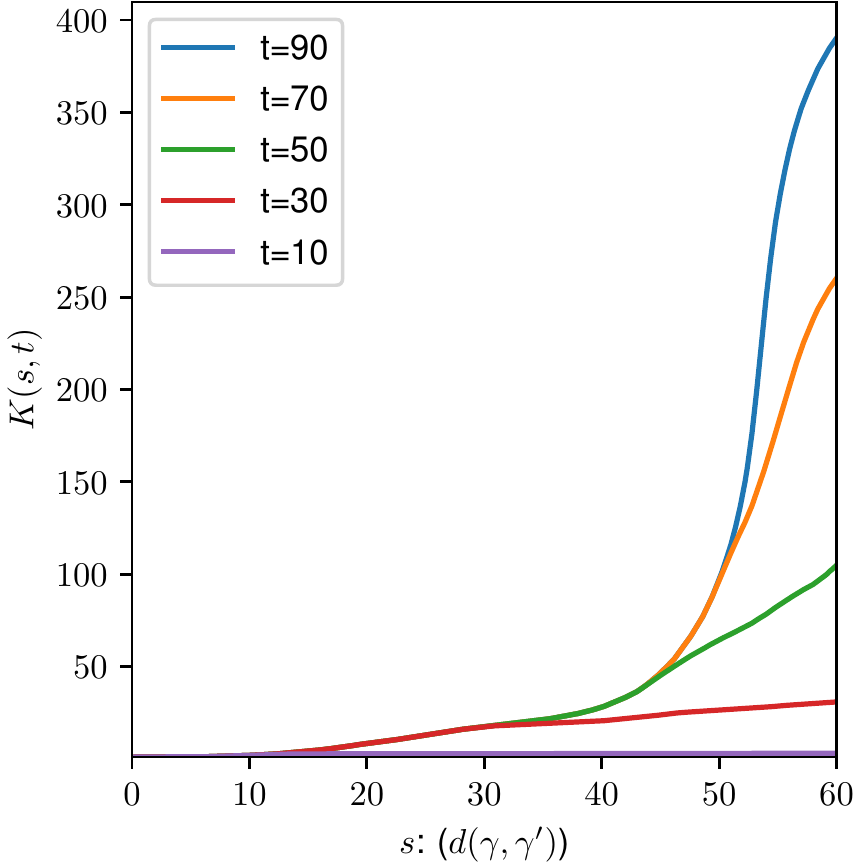}}
\\\includegraphics[width=0.23\textwidth, trim={ 9.85cm 0.2cm 1.8cm 1.2cm}, clip]{plots/K_uni_lin_randrot_v2_0.pdf}
&\includegraphics[width=0.23\textwidth, trim={ 9.85cm 0.2cm 1.8cm 1.2cm}, clip]{plots/K_uni_spiral_randrot_v2_0.pdf}
&\includegraphics[width=0.23\textwidth, trim={ 9.85cm 0.2cm 1.8cm 1.2cm}, clip]{plots/K_uni_brownian_v2_0.pdf}
&\includegraphics[width=0.23\textwidth, trim={ 9.85cm 0.2cm 1.8cm 1.2cm}, clip]{plots/K_Mog_linear_each_randrot_v2_0.pdf}
\end{tabular}
\caption{$K$-function on the generated data, where each column corresponds to one data set. (Top row): The data sets $X_1, X_2, X_3$ and $X_4$ described in 4.1. (Middle row): The $K$-function of the data set above for fixed values of $t$. (Bottom row): The $K$-function of the data set above for fixed values of $s$. }
\label{GeneratedDataset}
\end{figure}

The four generated data sets $X_1, X_2, X_3$ and $X_4$ each contain $500$ fibers with curve length $l = 40$ and center points in $[0,100]^3$ and is visualized in the first row of Fig. \ref{GeneratedDataset}. Each data set is created by sampling center points from a distribution on $\R^3$ and fibers from a distribution on $S_0$, that is then translated by the center points. For the first three data sets, the center points are generated by a Poisson process and the fibers are uniformly rotated lines in $X_1$, uniformly rotated spirals in $X_2$ and Brownian motions in $X_3$. The data set $X_4$ has clustered center points and within each cluster the fibers are slightly perturbed lines.  
\par
To avoid most edge effects, we choose the window $W \approx [13,87]^3\subset \R^3$ for the calculation of the $K$-function. Furthermore, we choose a Gaussian kernel $K_\sigma$ as in \eqref{gauskernel} with $p=2$ and $\sigma = \frac{100}{3}$. Finally, $c$ is defined to be the mass center of the curve.

The first row of Fig. \ref{GeneratedDataset} shows the generated data sets and the respective $K$-functions are visualized the second and third row. In the second row, $s\mapsto K(t,s)$ is plotted for fixed values of $t$. For example, the graphs with $t=50$ show the expected numbers of fibers within currents distance $s$, where the distance of center points are $50$ at most. Lastly in the third row, $t\mapsto K(t,s)$ is plotted for fixed values of $s$. Similarly, the graphs with $s=70$ show the expected numbers of fibers with center point distance $t$ when the currents distances are $70$ at most. Thus, the graphs in the second row capture the fiber shape difference of each data set whereas the graphs in the third row capture spatial difference.
\par
It is distribution $X_3$ that we consider to a natural suggestion for a uniform randomness distribution of fibers. This is because Brownian motions are well-known for modelling randomness, thus representing shape randomness. And by translating these Brownian motion with a Poisson process, we argue that this distribution is a good choice. 
\par 
Considering only the data sets with uniformly distributed center points, i.e., the first three columns of Fig. 2, we see a big difference in the second row of plots. This indicates that the $K$-function is sensitive to the change in shape. The $K$-function for the Brownian motions captures much more mass for smaller radii compared to the lines, with the spirals being somewhere in-between. The plots in the third row are very much as expected, since we generated the center points from a Poisson process. Finally, the second row plot for $X_4$ indicate a slight shape clustering when compared to the uniformly rotated lines. This makes sense, since each cluster is directed differently.  
\subsection{Application to myelin sheaths}
Myelin surrounds the nerve cell axons and is an example of a fiber structure in the brain. Based on 3D reconstructions from the region motor cortex of the mouse brain, centre lines were generated in the myelin sheaths. The data sets ST01, ST06, ST17 and ST20 displayed in the first row of Fig. 1 represent the myelin sheaths from four samples at different debts.
\par 
For real shape-valued data sets, it very common that only parts of the shapes are observed. This is the case for many fiber data sets as well. This fact is important to have in mind when choosing $c$, since we should have a clear idea of when $c(\gamma)$ is observed, in order to get an unbiased estimate. 
\par 
Since myelin sheaths tend to be quite long, we chose to divide the fibers of length greater than $40$ into several fibers segments of length $40$. This has the benefits, that the mass center is an appropriate choice for $c$ and that the results are comparable with the results of Fig. 2, since the curves are of similar length. 
\par 
The results of the estimated K-function on the four data sets ST01, ST06, ST17 and ST20 are visualized in Fig. 1, where $s\mapsto K(t,s)$ is plotted in the second row for fixed values of $t$ and $t\mapsto K(t,s)$ is plotted in the third row for fixed values of $s$. The plots in the second row showing the fiber shape are very similar, resembling the fiber distribution of $X_2$. We notice a slight difference in ST20, where the graphs have a more pronounced cut off.
When noticing the scale of the $y$-axis, we see that the expected number of neighbor fibers vary significantly between the data sets.  
\par 
The third row plots indicate that the center point distributions of each data set is similar to the center point distribution of $X_1$, $X_2$ and $X_3$, which we generated from a Poisson process. For ST17, we notice a slight clustering of center points for $t\in [10,15]$. The biggest difference is for the graphs for $t=30$, indicating that the neighbors for fibers i ST20 are of more similar shape than the others.  


\par 
\bibliographystyle{splncs04}
\bibliography{refs}

\end{document}